\documentclass[prc,superscriptaddress,unsortedaddress,twocolumn,showpacs,preprintnumbers,amsmath,amssymb]{revtex4}

\usepackage[dvipdfmx]{graphicx}
\usepackage{amsmath,amssymb,times}
\usepackage{color}
\usepackage{ulem}
\usepackage{bm}

\def\gsim{~\,\makebox(1,1){$\stackrel{>}{\widetilde{}}$}\,~}
\def\lsim{~\,\makebox(1,1){$\stackrel{<}{\widetilde{}}$}\,~}

\newcommand{\com}[1]{{\color[rgb]{0,0,1}{#1}}}

\newcommand{\beq}{\begin{equation}}
\newcommand{\eeq}{\end{equation}}
\newcommand{\bea}{\begin{eqnarray}}
\newcommand{\eea}{\end{eqnarray}}

\newcommand{\bfi}[1]{\mbox{\boldmath $#1$}}
\newcommand{\bfis}[1]{\mbox{\boldmath ${\scriptstyle #1}$}}
\newcommand{\vb}{{\bfi b}}
\newcommand{\vk}{{\bfi k}}

\newcommand{\vq}{{\bfi q}}

\newcommand{\vrr}{{\bfi r}}
\newcommand{\vR}{{\bfi R}}
\newcommand{\vQ}{{\bfi Q}}

\newcommand{\vib}{{\bfis b}}
\newcommand{\vik}{{\bfis k}}

\newcommand{\viR}{{\bfis R}}

\newcommand{\viq}{{\bfis q}}

\def\a{\alpha}

\def\ve{\varepsilon}

\begin{document}
\title{ 
Prediction  of the analyzing power for $\vec{p}$+$^{6}$He elastic scattering at 200~MeV \\
from $\vec{p}$+$^{4}$He elastic scattering at 200~MeV}

\author{Masahiro Ishii}
\email[]{ishii@phys.kyushu-u.ac.jp}
\affiliation{Department of Physics, Graduate School of Sciences, Kyushu University,
             Fukuoka 819-0395, Japan}   
             
\author{Yasunori Iseri}
\email[]{iseri@chiba-kc.ac.jp}
\affiliation{Chiba-Keizai College, Chiba 263-0021, Japan}

\author{Masanobu Yahiro}
\email[]{yahiro@phys.kyushu-u.ac.jp}
\affiliation{Department of Physics, Graduate School of Sciences, Kyushu University,
             Fukuoka 819-0395, Japan}             

\date{\today}

\begin{abstract}
We apply the cluster-folding (CF) model for $\vec{p}+^{6}$He 
scattering at 200~MeV, where the potential between 
$\vec{p}$ and $^{4}$He is fitted to data on  $\vec{p}+^{4}$He scattering at 200~MeV. 
For $\vec{p}+^{6}$He scattering at 200~MeV, the CF model reproduces measured differential cross section  
with no free parameter,  We then predict the analyzing power $A_y(q)$ with the CF model,  
where $q$ is the transfer momentum.  
Johnson, Al-Khalili and Tostevin  construct a theory for one-neutron halo scattering, 
taking (1) the adiabatic approximation 
and (2) neglecting the interaction between a valence neutron and a target, and 
yield a simple relationship between the elastic scattering of a
halo nucleus and of its core under certain conditions.
We improve their theory with (3) the eikonal approximation 
in order to determine $A_y(q)$ for  $^{6}$He from the data on $A_y(q)$ for  $^{4}$He.
The improved theory is accurate, when approximation (1)--(3) are good. 
Among the three approximations, approximation (2) is most essential. 
The CF model shows that approximation (2) is good in $0.9 \lsim q \lsim 2.4$~fm$^{-1}$. 
In the improved theory, the $A_y(q)$ for  $^{6}$He is the same as that for 
$^{4}$He. In $0.9 \lsim q \lsim 2.4$~fm$^{-1}$, we then predict  $A_y(q)$ for  $\vec{p}+^{6}$He scattering at 200~MeV  
from measured $A_y(q)$ for $\vec{p}+^{4}$He scattering at 200~MeV.  
We thus predict  $A_y(q)$ with the model-dependent and the model-independent prescription. 
The ratio of differential cross sections measured for $^{6}$He to that for $^{4}$He 
is related to the wave function of $^{6}$He. We then determine the radius between  $^{4}$He and 
the center-of-mass of valence two neutrons in  $^{6}$He. The radius is 5.77~fm.
\end{abstract}


\maketitle

\section{Introduction}
\label{Introduction}

In the framework of the shell model for nuclei, the central and 
spin-orbit potentials are important for understanding  nuclear structure. 
  The importance was first discovered by Mayer and Jensen. 
The central and spin-orbit potentials in various stable nuclei 
are similar to the real part of optical potential in 
the $\vec{p}$ elastic scattering   on 
the corresponding stable nuclei. 
The optical potentials  are 
well determined by measured differential cross sections 
$d\sigma/d\Omega$ and analyzing powers $A_y$.

In general, the central and spin-orbit potentials in the  scattering of 
unstable nuclei on a $\vec{p}$ target are different 
from  the case of stable nuclei, since unstable nuclei have larger radii 
than the stable nuclei with the common mass number~\cite{Toyokawa:2013uua,Watanabe:2014zea}.

For scattering of $^{6}$He on a $\vec{p}$ target at an incident energy 
$E_{\rm lab}=71$~MeV,  the  $A_y$ was obtained in the inverse measurement~\cite{Hatano2005,Uesaka:2010mm,Sakaguchi:2011rp}. In the experiment, the $d\sigma/d\Omega$ is measured 
in $1.1 \lsim q \lsim 2.2$~fm$^{-1}$ ($42^\circ \lsim \theta_{\rm cm} \lsim 87^\circ$) and the  $A_y$ is in 
$1.0 \lsim q \lsim 1.9$~fm$^{-1}$ ($37^\circ \lsim \theta_{\rm cm} \lsim 74^\circ$)~\cite{Hatano2005,Uesaka:2010mm,Sakaguchi:2011rp}, where $q$  and $\theta_{\rm cm}$ are the transfer momentum and the scattering angle 
 in the center-of-mass frame, respectively.  
The measured $A_y$ is reproduced by the the cluster-folding (CF) model~\cite{Sakaguchi:2011rp}. 
It is shown in Ref.~\cite{Sakaguchi:2011rp}  that 
the spin-orbit part of the phenomenological optical potential has a shallow and long-ranged shape.  
This problem is not solved yet.

The same measurement was made for $E_{\rm lab} = 200$~MeV~\cite{Chebotaryov:2018ilv}, since 
the nucleon-nucleon (NN) total cross section has a minimum around there. 
However, the result was shown only for $d\sigma/d\Omega$ in  
$1.7 \lsim q \lsim 2.7$~fm$^{-1}$ ($36^\circ \lsim \theta_{\rm cm} \lsim 59^\circ$).

The $\vec{p}+^{4,6}$He scattering at $E_{\rm lab}=200$~MeV were analyzed by the Melbourne g-matrix 
folding model~\cite{Toyokawa:2013uua}. 
The model predicted $d\sigma/d\Omega$  and $A_y$ for $^{6}$He, but 
not does account for the data~\cite{Moss1980} for $^{4}$He 
in  $ q \gsim 3.3$~fm$^{-1}$ ($\theta_{\rm cm} \gsim 80^\circ$).
{\it Ab initio} folding potentials based on no-core shell-model~\cite{Burrows:2018ggt} 
were  constructed and applied for  $\vec{p}$+$^{4,6}$He scattering at $E_{\rm lab}=200$~MeV. 
The model reproduces the data on $d\sigma/d\Omega$ for $^{6}$He, but 
not $d\sigma/d\Omega$ for $^{4}$He in  $ q \gsim 2.5$~fm$^{-1}$ ($\theta_{\rm cm} \gsim 60^\circ$). 

Crespo and Moro calculated $d\sigma/d\Omega$ and $A_y$ for the $\vec{p}+^{4,6,8}$He scattering 
at $E_{\rm lab}=297$~MeV, using the Multiple Scattering expansion~\cite{Crespo:2007zz}.  
Microscopic optical potentials derived from NN $t$ matrix and nonlocal density was 
applied to the $\vec{p}+^{4}$He scattering at $E_{\rm lab}=200$~MeV~\cite{Gennari:2017yez}, and reproduced 
the data of Ref.~\cite{Moss1980} in $ q \lsim 4.1$~fm$^{-1}$ ($\theta_{\rm cm} \lsim 110^\circ$). 

Johnson, Al-Khalili and Tostevin constructed a theory, using the adiabatic approximation 
and neglecting the interaction between a valence neutron and a target 
for one-neutron halo scattering~\cite{Johnson:1997zz}. 
They yield a simple relationship between the elastic scattering of a
halo nucleus and of its core from a stable target. The relation is good, 
if (1) the adiabatic approximation is accurate and (2) the potential between a valence neutron 
and a target can be switched off.  
In the present paper, we refer to the theory of Ref.~~\cite{Johnson:1997zz} as valence-core cutting (VCC) 
theory.    
When the VCC theory is applied to $\vec{p}+^{6}$He scattering, 
the relation is Eq.~\eqref{Eq,JAT} in Sec.~\ref{Sec,Four-body model}.

In this paper, we improve the VCC theory for $\vec{p}+^{6}$He scattering at $E_{\rm lab}=71$ and $200$~MeV, using (3) the eikonal approximation in addition to approximations (1) and (2).  Among  the approximations, approximation (2) is most essential and should be investigated.   
Using the CF mode, we confirm that approximation (2)  is good 
in $0.9 \lsim q \lsim 2.4$~fm$^{-1}$ for $E_{\rm lab}=200$~MeV, but good only 
in the vicinity of $q =0.9$~fm$^{-1}$ for $E_{\rm lab}=71$~MeV. 

In the improved VCC theory, the $A_y$ for  $^{6}$He is the same as that for 
$^{4}$He.  In $0.9 \lsim q \lsim 2.4$~fm$^{-1}$, we can predict  
$A_y(q)$ for  $\vec{p}+^{6}$He scattering at 200~MeV from  $A_y(q)$ 
measured for $\vec{p}+^{4}$He scattering at 200~MeV without using any model.  
Since the ratio of $d\sigma/d\Omega$ for $^{6}$He to that for $^{4}$He is related 
to the wave function of $^{6}$He, we can determine the radius between  $^{6}$He and 
the center-of-mass of valence two neutrons from the ratio.

In order test to approximation (2), we use the CF model for $\vec{p}+^{6}$He 
scattering at 200~MeV, where the potential between 
$\vec{p}$ and $^{4}$He is fitted to data on $\vec{p}+^{4}$He scattering at 200~MeV. 
The CF model reproduces the differential cross section for $\vec{p}+^{6}$He 
scattering with no free parameter. We then predict  $A_y$.

The improved VCC theory and the results are shown in Sec.~\ref{Sec,Four-body model}. 
The CF model is explained and its results are shown in Sec.~\ref{sec:Folding model}.
Section \ref{Summary} is devoted to a summary. 


\section{Improved VCC theory and its results}
\label{Sec,Four-body model}

\begin{figure}[tb]
\includegraphics[width=0.70\linewidth,clip]{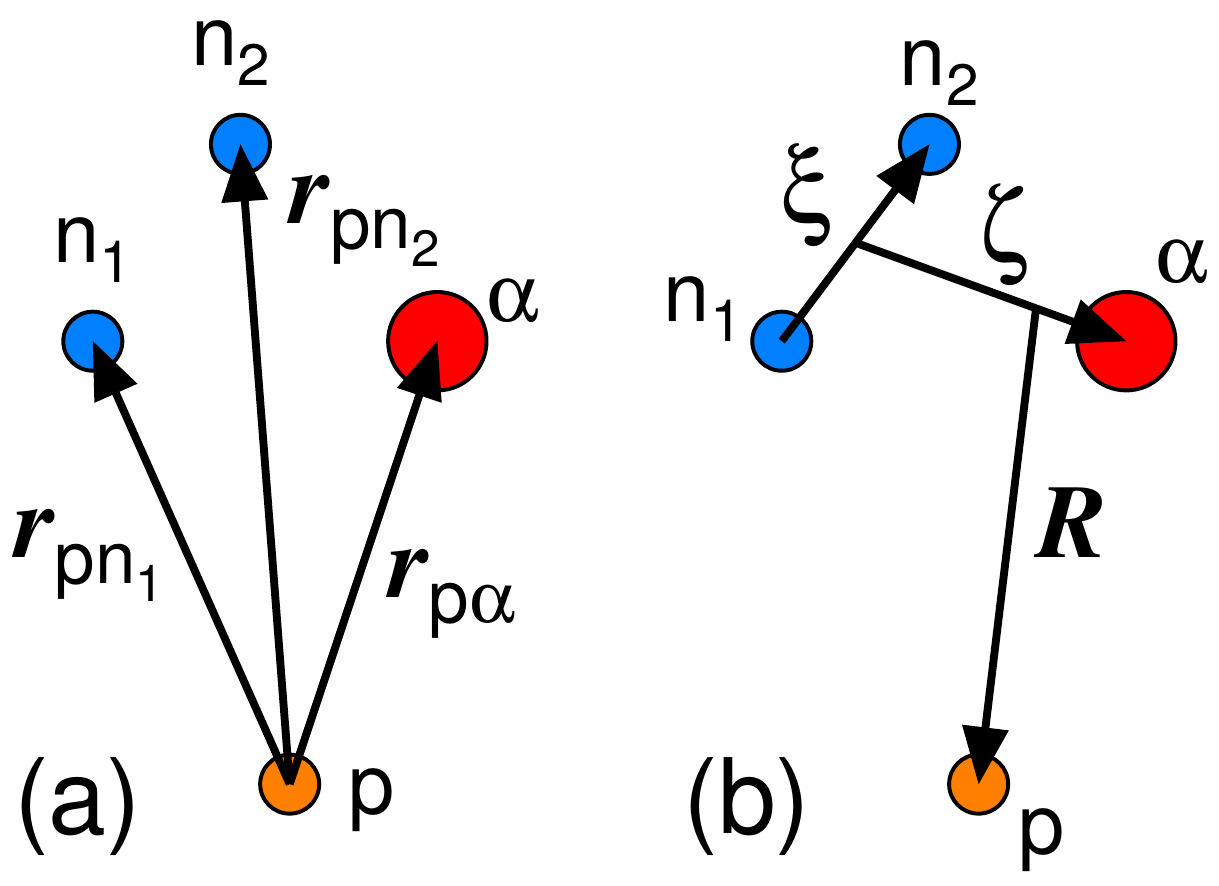}
\caption{
Two sets of coordinates in four-body model.
}
\label{fig:coord}
 \end{figure}

We start with the $p$+$n_1$+$n_2$+$^{4}$He four-body model  to 
consider the $\vec{p}$ elastic scattering on $^{6}$He  at $E_{\rm lab}=$71 and $200$~MeV;
see Fig.~\ref{fig:coord} for two coordinate sets of the four-body system.
The total Hamiltonian of the scattering is 
\begin{align}
 H &=-\frac{\hbar^2}{2\mu_6}\nabla_{R}^2+U+H_{{6}},\\
 U &=U_{pn_1}({r}_{pn_1})+U_{pn_2}({r}_{pn_2})+U_{p\alpha}({r}_{p\alpha} ) +V_{p\alpha}^{\rm Coul}({r}_{p\alpha})
\end{align}
where  $\mu_6$ is the reduce mass between $\vec{p}$ and $^{6}$He
and the Hamiltonian $H_{6}$ of $^{6}$He is 
described by the $n_1$+$n_2$+$^{4}$He three-body model. 
The coordinates $\bm{r}_{p\gamma}$ for $\gamma=n_1,n_2, \alpha$ are shown in Fig.~\ref{fig:coord} (a).
The $U_{p\gamma}$ are the nuclear interaction between $\vec{p}$ and $\gamma$.

The exact $T$ - matrix of the elastic scattering is
\bea
T & = &\langle e^{i\vik \cdot \viR}\Phi |\ U |\ \Psi \rangle
\eea 
for the total wave function $\Psi$, the incident momentum $\vk$. 
The ground state $\Phi$ of  $^{6}$He has an energy $\ve_0$.

Following Ref.~\cite{Johnson:1997zz}, we take the adiabatic approximation 
to the total wave function $\Psi$ and neglect the interactions $U_{pn_1}$ and $U_{pn_2}$. 
The resulting Hamiltonian is 
\begin{align}
 H_{\rm AD} &=-\frac{\hbar^2}{2\mu_6}\nabla_R^2+U_{p\alpha}({r}_{p\alpha} )+
 V_{p\alpha}^{\rm Coul}({r}_{p\alpha})+\ve_0,\\
    &=-\frac{\hbar^2}{2\mu_6}\nabla_{r_{p\a}}^2+U_{p\alpha}({r}_{p\alpha} )
    +V_{p\alpha}^{\rm Coul}({r}_{p\alpha})+\ve_0, 
\end{align}
where $\nabla_R^2=\nabla_{r_{p\a}}^2$ as a result of the transform from $\vR$ to $\vrr_{p\a}$. 
The initial wave function of $\Psi$ is 
\bea
e^{i\vik \cdot \viR}\Phi =e^{i\vik \cdot ({\vrr}_{p\alpha}-\a_{vc}\zeta)}\Phi  
\eea
with $\a_{vc}=2 /(4+2)=1/3$. We then obtain  
\bea
T_{\rm AD} & = &\langle e^{i\vik \cdot \viR}\Phi |\ U_{p\alpha}+V_{p\alpha}^{\rm Coul}({r}_{p\alpha}) | e^{-i  \a_{vc} \vik \cdot \zeta}\Phi \chi_{\vik}({r}_{p\alpha} ) \rangle
\notag \\
\eea 
with the distorting wave function  $\chi_{\vk}({\vrr}_{p\alpha} )$ defined by 
\bea
&&\chi_{\vk}({\vrr}_{p\alpha} ) 
\notag \\
&=&
\frac{i \varepsilon}{E_{\rm cm}-\frac{\hbar^2}{2\mu_6}\nabla_{\vrr_{p\a}}^2+U_{p\alpha}({r}_{p\alpha} )+V_{p\alpha}^{\rm Coul}({r}_{p\alpha})+i \varepsilon} 
e^{i\vik \cdot {\vrr}_{p\alpha}}  
\notag \\
\eea
with infinitesimally small $\varepsilon$ and  the incident energy $E_{\rm cm}=\hbar^2 k^2/(2\mu_6)$ 
in the center of mass system.  
The $\chi_{\vk}({\vrr}_{p\alpha} )$ is the distorting wave function between $\vec{p}$ and $^{4}$He 
with the reduced mass $\mu_6$, and not the distorting wave function 
of the $\vec{p}$+$^{4}$He elastic scattering with the same incident energy 
$E_{\rm lab}$, because the reduced mass $\mu_6$ between $\vec{p}$ and $^{6}$He is different from the reduced mass $\mu_4$ between  $\vec{p}$ and $^{4}$He.

The $T_{\rm AD}$ becomes  
\bea
T_{\rm AD}  = 
F( \a_{vc} (\vik - \vik') )
\langle e^{i\vik' \cdot {\vrr}_{p\alpha}} | U_{p\alpha}+ V_{p\alpha}^{\rm Coul}| \chi_{\vik}({\vrr}_{p\alpha} ) \rangle_{{\vrr}_{p\alpha}}
\label{eq:T-ad}
\eea 
with the form factor 
\bea
F(\vQ) \equiv F( \a_{vc} (\vik - \vik') ) = \langle e^{i  \a_{vc} (\vik - \vik') \cdot \zeta} |\Phi |^2 \rangle_{\zeta \xi}  
\eea
for $\vQ \equiv \a_{vc} (\vik - \vik') = \vq/3$.  

Using Eq.~\eqref{eq:T-ad}, we can get the differential cross section as 
\bea
\Big( \frac{d \sigma}{d \Omega} \Big)_{p+^{6} {\rm He}}
=| F( \vQ ) |^2 \Big( \frac{d \sigma}{d \Omega} \Big)_{p+^{4} {\rm He}}^{\mu_6} .
\label{Eq,JAT}
\eea
This equation was derived in Ref.~\cite{Johnson:1997zz}. 
In the right hand side of  Eq.~\eqref{Eq,JAT},  the part 
$\Big( \frac{d \sigma}{d \Omega} \Big)_{p+^{4} {\rm He}}^{\mu_6}$ 
is calculated theoretically~\cite{Johnson:1997zz}.

Now we improve Eq.~\eqref{Eq,JAT} in order to determine $| F( \vQ ) |$ from 
experimental data  on  $\vec{p}$ + $^{4,6}$He scattering at the same $E_{\rm lab}$. 
The incident energy $E_{\rm lab}$ in the laboratory system is determined by 
the velocity $v$ as 
\bea
E_{\rm lab}=\frac{v^2M_{p}}{2}
\label{Eq:E-v}
\eea
for proton mass $M_{p}$.
When we apply the eikonal approximation to the p+$^{4}${\rm He} scattering, the scattering amplitude is 
\bea
&&f_{p\a} = \frac{i \mu_4 v}{2\pi \hbar}
\int d\vb\;
e^{- i \viq \cdot \vib}
(1-e^{i \chi(\vib)}) \;
\label{eq:f}
\eea
with
\bea
\chi(\vb)
&=&-{1 \over \hbar v} \int_{-\infty}^{\infty}
dz \; [U_{p\alpha}(z,\vb) + V_{p\alpha}^{\rm Coul}(z,\vb) ] 
\label{eq:chi}
\eea
for $\vrr_{p\a}=(\vb,z)$. The differential cross section is thus determined by $v$, i.e., $E_{\rm lab}$. 
We then obtain 
\bea
\Big( \frac{d \sigma}{d \Omega} \Big)_{p+^{6} {\rm He}}^{E_{\rm lab}}
=| F( \vQ ) |^2 \Big( \frac{d \sigma}{d \Omega} \Big)_{p+^{4} {\rm He}}^{E_{\rm lab}}
\Big( \frac{\mu_6}{\mu_4} \Big)^2 .
\label{eq:Xsec-eikonal}
\eea
from Eqs.~\eqref{eq:f}-\eqref{eq:chi}. 
The equation \eqref{eq:Xsec-eikonal} allows us to determine $| F( \vQ ) |$ from 
two  differential cross sections measured for $\vec{p}$ + $^{4}$He and  $\vec{p}$ + $^{6}$He scattering 
at a common $E_{\rm lab}$. 

When $p$ is polarized, 
the factor $(| F( \vQ ) |{\mu_6}/{\mu_4})^2$ is common between 
the cross section for incident proton having up-spin  and that for proton having down-spin. This means 
that the vector analyzing $A_y(q)$ for $\vec{p}$ + $^{6}$He scattering is 
the same as $A_y(q)$ for $\vec{p}$+ $^{4}$He in the improved VCC theory.  

The relation \eqref{eq:Xsec-eikonal} between 
$\Big( \frac{d \sigma}{d \Omega} \Big)_{p+^{6} {\rm He}}^{E_{\rm lab}}$ 
and $\Big( \frac{d \sigma}{d \Omega} \Big)_{p+^{4} {\rm He}}^{E_{\rm lab}}$ is good, 
when the eikonal and adiabatic approximations are good and $U_{pn_1}=U_{pn_2}=0$.  
It is shown in Ref.~~\cite{Yahiro:2008dr} that the eikonal and adiabatic approximations are good 
for a few hundred MeV. The approximation $U_{pn_1}=U_{pn_2}=0$ is good in $0.9 \lsim q \lsim 2.4$~fm$^{-1}$ 
for 200~MeV as shown in Sec. \ref{Sec,Xsec,Ay-200}, but good only near $q =0.9$~fm$^{-1}$  for 71 MeV 
as mentioned in Sec.~\ref{Sec,Xsec,Ay-71}.


\subsection{Determination of $|F|$ from measured differential cross sections for $\vec{p}$ + $^{4,6}$He scattering}
\label{sec:The form factor}

Using Eq.~\eqref{eq:Xsec-eikonal}, we can determine $|F(Q)|$ from experimental data on the cross sections of 
$p$+ $^{4,6}$He scattering at the same $E_{\rm lab}$, when the most essential condition 
$U_{pn_1}=U_{pn_2}=0$ is good and the angular momentum between $n_1$ and $n_2$  is zero.

\begin{figure}[htbp]
 \begin{center}
  \includegraphics[width=0.9\linewidth]{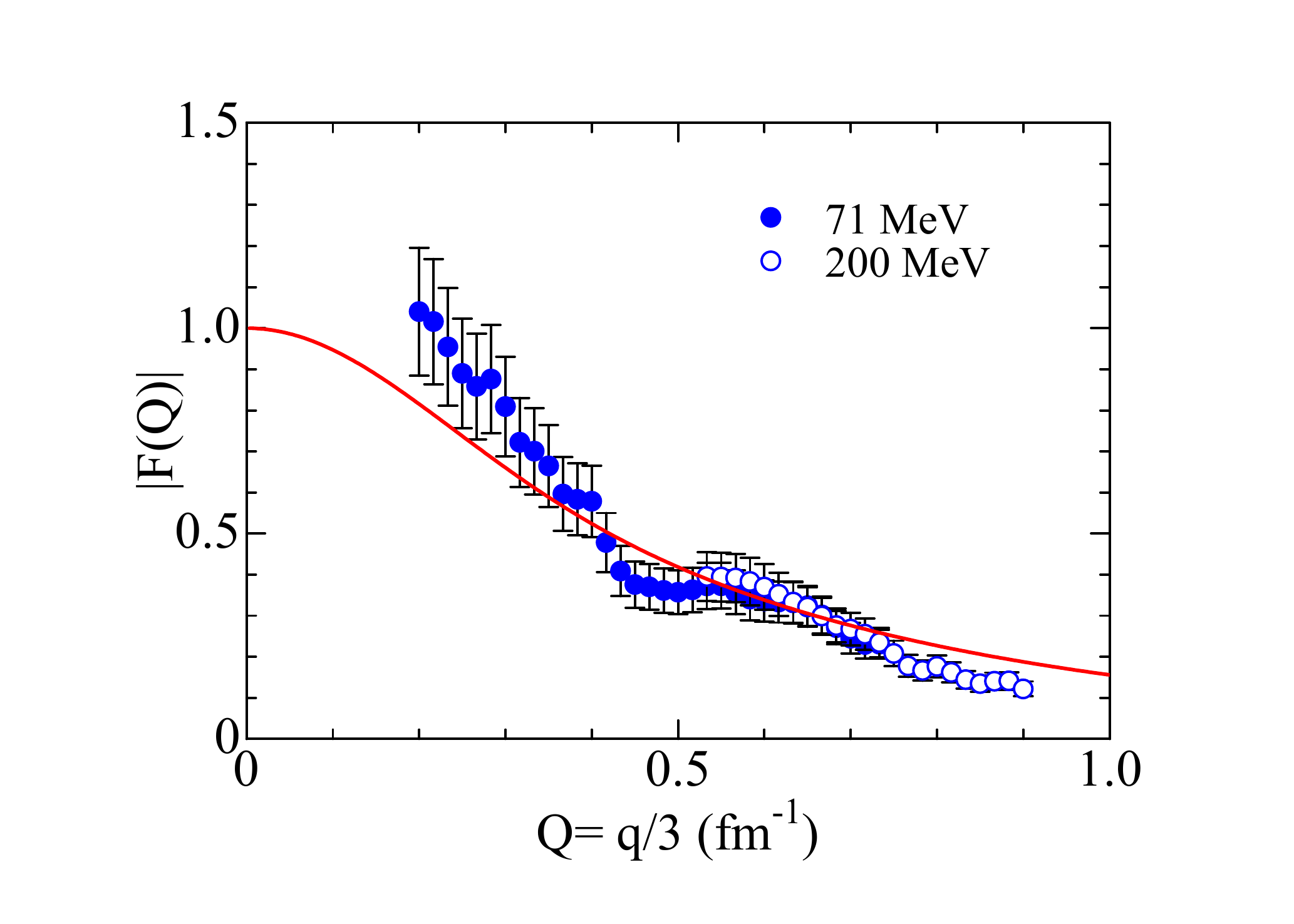}
  \caption{$Q$ dependence of $|F|$. 
  The solid (open) circles are the result determined from the experimental data at 71 (200) MeV.
  The solid line is a result of $V(\zeta)=-25\exp[-(\zeta/1.41)^2]$. 
    Experimental data are taken from Refs.~\cite{Burzynski:1989zz,Korsheninnikov:1997qta,Sakaguchi:2011rp} for 71~MeV and Refs.~\cite{Moss1980,Chebotaryov:2018ilv} for 200~MeV.
}
  \label{fig:F-Q-exp-th.pdf}
 \end{center}
\end{figure}

As for $E_{\rm lab}=200$~MeV, the data are available in Ref.~\cite{Moss1980} 
for $^{4}$He and in Ref.~\cite{Chebotaryov:2018ilv} for $^{6}$He.
As for $E_{\rm lab} =71$~MeV, the data are available 
in Refs.~\cite{Sakaguchi:2011rp,Korsheninnikov:1997qta} for
$^{6}$He, but not for $^{4}$He. We then take the data~\cite{Burzynski:1989zz}  
on $\vec{p}$ + $^{4}$He scattering at $E_{\rm lab}=72$~MeV. 
The resulting  $|F(Q)|$ is smooth, as shown in Fig. \ref{fig:F-Q-exp-th.pdf}.  
The approximation $U_{pn_1}=U_{pn_2}=0$ is good in $0.9 \lsim q \lsim 2.4$~fm$^{-1}$ 
for 200~MeV as shown in Sec. \ref{Sec,Xsec,Ay-200},  but good only in the vicinity of $q =0.9$~fm$^{-1}$  
for 71 MeV as mentioned in  \ref{Sec,Xsec,Ay-71}. In Fig. \ref{fig:F-Q-exp-th.pdf}, the resulting $|F(Q)|$ is thus reliable in   $0.3 \lsim Q \lsim 0.8$~fm$^{-1}$.

The Fourier transform $|F(\zeta)|$ of $|F(Q)|$ is a function of $\zeta$. 
We then assume that the potential between  $^{4}$He and the center-of-mass of 
$n_1$ and $n_2$ is a one-range Gauss function $V(\zeta)$, and can obtain  $|F(Q)|$ by solving 
Schrodinger equation with the potential. 
The solid line denotes a result of $V(\zeta)=-25\exp[-(\zeta/1.41)^2]$, and reproduces the experimental 
$|F(Q)|$ for 200~MeV. The resulting radius between  $^{4}$He and the center-of-mass of $n_1$ and $n_2$ is 5.77~fm. The corresponding binding energy is 0.172~MeV.

\subsection{Model independent prediction on $A_y$ for $\vec{p}$ + $^{4,6}$He scattering at 200~MeV}
\label{sec:The form factor}

When $p$ is polarized, 
the factor $| F( \a_{vc} (\vik - \vik') ) |{\mu_6}/{\mu_4}$ is common between 
the cross section for incident proton having up-spin  and that for proton having down-spin. This means 
that the vector analyzing $A_y(q)$ for $\vec{p}$ + $^{6}$He scattering is 
the same as $A_y(q)$ for $\vec{p}$+ $^{4}$He, when the condition $U_{pn_1}=U_{pn_2}=0$ is good. 
As mentioned later in Sec.~\ref{Sec,Xsec,Ay-200}, the condition is well satisfied 
in  $0.9 \lsim q \lsim 2.4$~fm$^{-1}$. 

We make a model-independent prediction on  $A_y(q)$ for $^{6}$He, 
assuming that the $A_y(q)$ for $^{6}$He is the same as the measured $A_y(q)$ of Ref.~\cite{Moss1980} for $^{4}$He. 
The predicted  $A_y(q)$ can be transformed into $A_y(\theta)$. 

\begin{figure}[htbp]
 \begin{center}
  \includegraphics[width=0.9\linewidth]{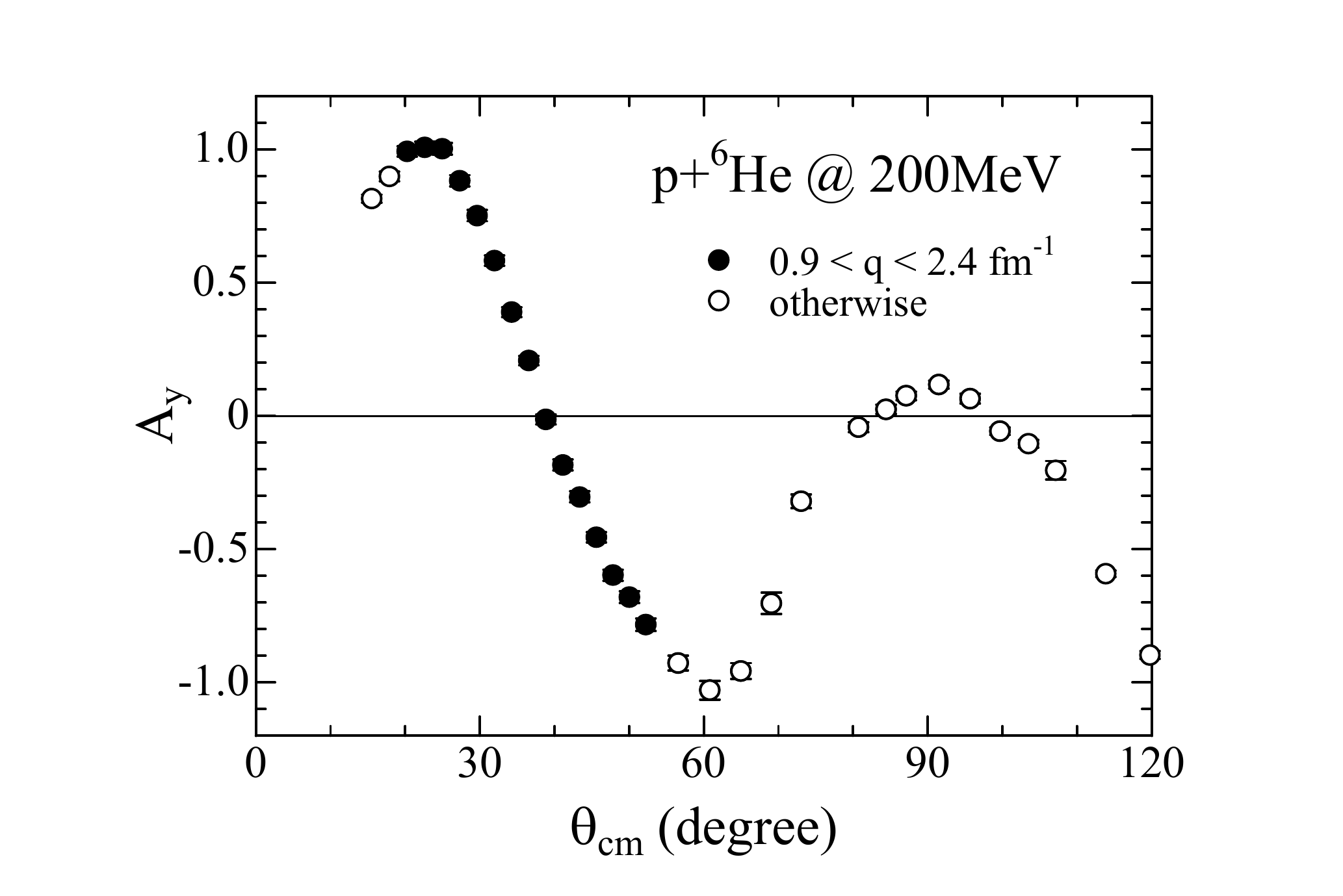}
  \caption{ $\theta$ dependence of predicted $A_y$ for  $\vec{p}$+ $^{6}$He scattering at 200~MeV 
See the text for closed and open circles. 
}
  \label{fig:Ay-exp}
 \end{center}
\end{figure}

Figure~\ref{fig:Ay-exp} shows the predicted $A_y(\theta)$ for $^{6}$He.   
The predicted $A_y(\theta)$  is reliable in $20^\circ \lsim \theta_{\rm cm} \lsim 55^\circ$ 
($0.9 \lsim q \lsim 2.4$~fm$^{-1}$).  
The reliable prediction in $20^\circ \lsim \theta_{\rm cm} \lsim 55^\circ$ are denoted 
by closed circles. It should be noted that our prediction shown by open circles are not good.

\subsection{$A_y$ for 71~MeV} 
\label{Sec,Ay-71}

\begin{figure}[htbp]
 \begin{center}
  \includegraphics[width=0.9\linewidth]{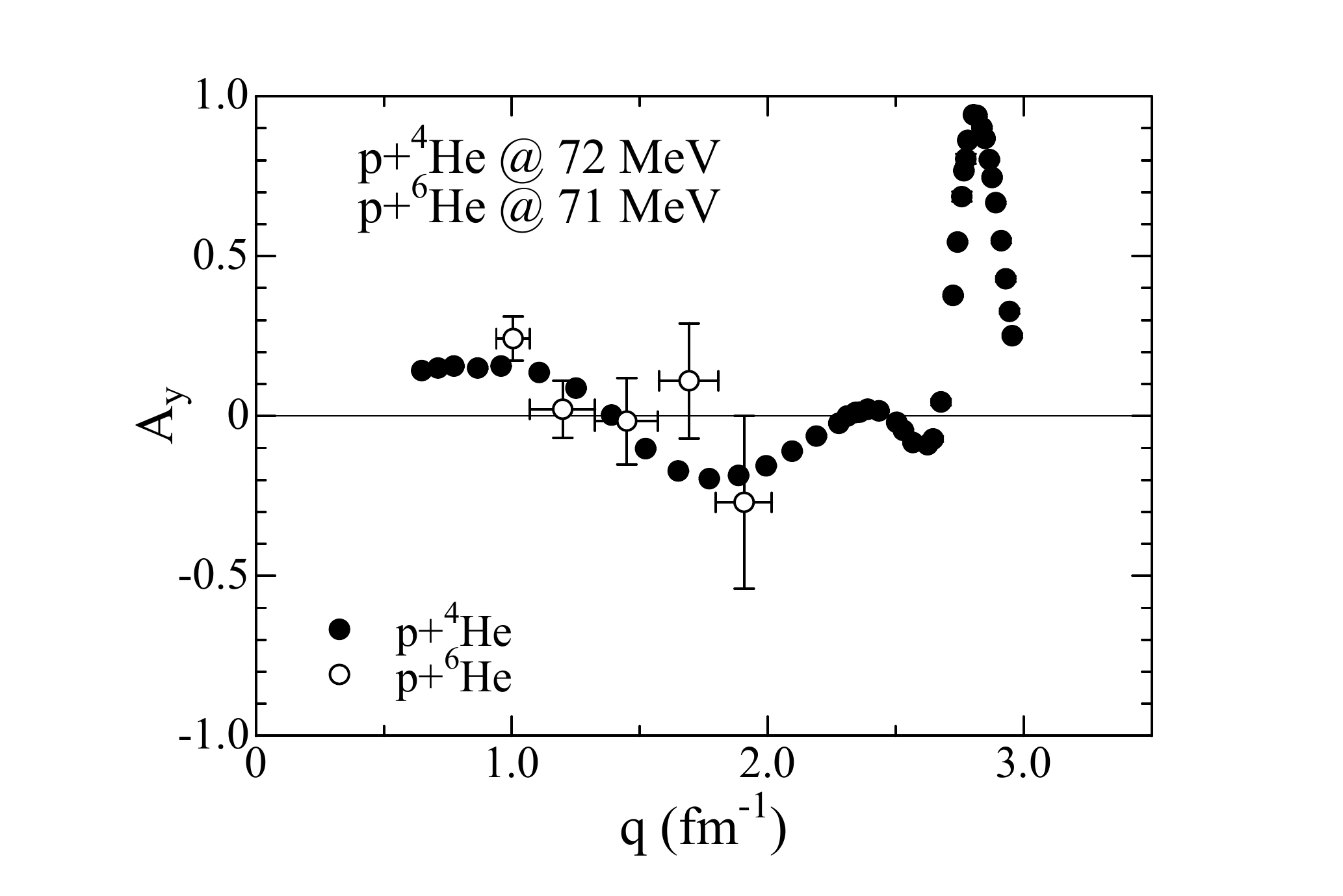}
  \caption{
$q$ dependence of measured $A_y$ (closed circles) for $\vec{p}$+$^{4}$He scattering at $E_{\rm lab} = 72$~MeV and 
measured $A_y$ (open circles) for $\vec{p}$+$^{6}$He scattering at $E_{\rm lab} = 71$~MeV.
 Data are taken from Ref.~\cite{Sakaguchi:2011rp} for $^{6}$He and Ref.~\cite{Burzynski:1989zz} for $^{4}$He. 
  }
  \label{fig:Ay-72}
 \end{center}
\end{figure}

Figure \ref{fig:Ay-72} shows $q$ dependence of $A_y$  measured for $\vec{p}$+$^{4}$He scattering at $E_{\rm lab} = 72$~MeV and that for $\vec{p}$+$^{6}$He scattering at $E_{\rm lab} = 71$~MeV. 
The $A_y$  for $^{6}$He is close to that for $^{4}$He, except for a data at $q=1.71$~fm$^{-1}$. 
The property can be analyzed quantitatively by the Jensen-Shannon (JS) divergence~\cite{lin1991divergence}.  
We show the analysis in Appendix \ref{Sec,JSD-Ay-71},  since the analysis is  new but has recently been used 
by LIGO Scientific and Virgo Collaborations~\cite{LIGOScientific:2018mvr}. 

\section{Cluster-folding model}
\label{sec:Folding model}

We  consider the cluster folding (CF) model  for 
the $\vec{p}$ elastic scattering from $^{6}$He at $E_{\rm lab} =200$~MeV. 
In addition, we recalculate  the $\vec{p}$ elastic scattering from $^{6}$He at $E_{\rm lab} =71$~MeV in order to obtain the $F$.

Following Ref.~\cite{Sakaguchi:2011rp}, we derive 
the nuclear potential $U_{\textrm{CF}}({R})$ between $\vec{p}$ and $^{6}$He with 
the $^{6}$He density~\cite{Hiyama96,Hiyama03} obtained by $\alpha n n$ OCM:
\bea
U_{\textrm{CF}}({R}) 
&=& \int U_{p n_1} \, \rho_n^{\textrm{CF}} (r_1) \, d\bm{r}_1
      + \int U_{p n_2} \, \rho_n^{\textrm{CF}} (r_2) \, d\bm{r}_2\nonumber \\
      &+& \int U_{p \alpha} \, \rho_\alpha^{\textrm{CF}} (r_\alpha) \, d\bm{r}_\alpha , 
\label{eq:ucf1}
\eea
with 
\bea
U_{pn_1} &=& U^0_{pn}({r}_{pn_1}) +
U^\text{LS}_{pn}({r}_{pn_1})
\bm{\ell}_{pn_1}\cdot(\bm{\sigma}_p+\bm{\sigma}_{n_1}), 
\\
U_{pn_2} &=& U^0_{pn}({r}_{pn_2}) +
U^\text{LS}_{pn}({r}_{pn_2})
\bm{\ell}_{pn_2}\cdot(\bm{\sigma}_p+\bm{\sigma}_{n_2}), 
\\
U_{p\alpha}&=&U^0_{p\alpha}({r}_{p\alpha} ) +
U^\text{LS}_{p\alpha}({r}_{p\alpha} ) 
 \bm{\ell}_{p\alpha} \cdot \bm{\sigma}_{p}~,~~~~~~~~~~~~~~~~~~~~~~~
\label{eq:vp}
\eea
where the coordinates $\vrr_1$, $\vrr_2$ and $\vrr_{\a}$  are the position vectors of 
$n_1$, $n_2$, and the alpha core from the center of mass of $^{6}$He, respectively, and 
$\rho_n^{\textrm{CF}}$ and $\rho_\alpha^{\textrm{CF}}$ are the neutron and $\alpha$ densities, respectively.

We can rewrite the $U_{\textrm{CF}}(R)$ into
\begin{equation}
U_{\textrm{CF}} = U_0^{\textrm{CF}}(R) + U_{\textrm{LS}}^{\textrm{CF}}(R) \, \bm{L} \cdot \bm{\sigma}_p  \;,
\label{eq:ucf2}
\end{equation}
with the central part 
\begin{eqnarray}
U_0^{\textrm{CF}}(R) &=& 2 \, \int U_{pn}^0 (|\bm{r}_1 - \bm{R}| ) \, \rho _n^{\textrm{CF}} (r_1) \, d \bm{r}_1\nonumber\\
  &+& \int U_{p \alpha}^0 (|\bm{r}_\alpha - \bm{R}| ) \, \rho_\alpha^{\textrm{CF}} (r_\alpha) \, d \bm{r}_\alpha
\label{eq:ucf0}
\end{eqnarray}
and the spin-orbit part
\begin{eqnarray}
U_{\textrm{LS}}^{\textrm{CF}}(R) &=& \frac{1}{3} \, \int U_{pn}^{\textrm{LS}} (|\bm{r}_1 - \bm{R}| ) \, 
           \left\{ 1 - \frac{\bm{r}_1 \cdot \bm{R}}{R^2} \right\} \, \rho _n^{\textrm{CF}} (r_1) \, d \bm{r}_1
               \nonumber \\
  &+& \frac{2}{3} \, \int U_{p \alpha}^{\textrm{LS}} (|\bm{r}_\alpha - \bm{R}| ) \, 
           \left\{ 1 - \frac{\bm{r}_\alpha \cdot \bm{R}}{R^2} \right\} \,
           \rho_\alpha^{\textrm{CF}} (r_\alpha) \, d \bm{r}_\alpha .
               \nonumber \\
\label{eq:ucfls}
\end{eqnarray}

In the derivation of Eq.~\eqref{eq:ucfls}, the following points have been used;
(I) the internal momenta  of ${}^4$He and their expectation values are effectively zero for a spherically symmetric nucleus, and (II) the  internal coordinates  contribute to $\bm{L}$ by its component 
along the $\bm{R}$ direction. 
Eventually we have used  
\begin{equation}
\bm{\ell}_{p \gamma} = C_{\gamma} \, \bm{L} \, ( 1 - \frac{\bm{r}_{\gamma} 
\cdot \bm{R}}{R^2} ) \; 
\label{eq:ell3}
\end{equation}
with $C_{\gamma}=1/6$ for $\gamma=n_1,n_2$ and $C_{\gamma}=2/3$ 
for $\gamma=\alpha$. 

The $U_{p\a}$ is the optical potential (OP), 
and $U_{pn_1}$ and $U_{pn_1}$ are the CEG~\cite{Yamaguchi83,Nagata85,Yamaguchi86}. 
The $g$ matrix, derived from the Hamada-Johnston potential~\cite{Hamada62}, is successful 
in reproducing the data on  $\vec{p}$ elastic scattering 
from many nuclei in a wide range of incident energies, $E_{\rm lab}=$ 20--200~MeV~\cite{Yamaguchi83,Nagata85,Yamaguchi86}. 
For $\vec{p}$ + $^{6}$He elastic scattering at 71~MeV, the CF model well 
reproduces the data on differential cross sections and $A_y$~\cite{Sakaguchi:2011rp}.

\subsection{Potential fitting of $\vec{p}$+$^4$He scattering and results of CF model 
$\vec{p}$+$^6$He scattering}

We now fit the OP potential $U_{p \alpha}$ to data~\cite{Moss1980} for $\vec{p}$+$^4$He scattering at $E_{\rm lab}=200$~MeV with a  Woods-Saxon form:
\begin{eqnarray}
U_{p \alpha} =  &-&V_0 \, f_r (r_{p{\alpha}}) - i \, W_0 \, f_i (r_{p{\alpha}})  
\nonumber  \\
                      &+& 4i\, a_{id} \, W_{id} \, \frac{d}{dr_{p{\alpha}}} f_{id}(r_{p{\alpha}})  
\nonumber \\
      &+&  V_s \, \frac{2}{r_{p{\alpha}}} \, \frac{d}{dR} f_s (r_{p{\alpha}}) \; \bm{\ell}_{p\alpha} \cdot \bm{\sigma}_p
\label{eq:op1}
\end{eqnarray}
with
\begin{eqnarray}
f_x (r_{p{\alpha}}) &=& \left[ 1 + \exp \left( \frac{r_{p{\alpha}}-r_{x} A^{1/3}}{a_x}
                        \right) \right] ^{-1}  
\label{eq:op2}
\end{eqnarray}
for $x=r,i,id, s$, where 
$\bm{\sigma}_{p}$ stands for the Pauli
spin operator of an incident proton. 
The Coulomb potential between the proton and $^{4}$He ($^{6}$He) is obtained from the
uniformly charged sphere with the radius $1.4A^{1/3}$, where $A=4$ for $^{4}$He and $A=6$ for $^{6}$He.

The best-fit potential parameters are obtained  by minimizing the
$\chi^2$ values of $d\sigma/d\Omega$ and $A_y$. 
The resulting parameter set is tabulated in Table~\ref{table:param}, together with the case of 
$E_{\rm lab}=72$~MeV of Ref.~\cite{Sakaguchi:2011rp}. 

First of all, we briefly shows  results of the OP and the CF model in Fig.~\ref{fig:He4,6-200}.  
The left panel shows that our fitting is good for $\vec{p}$+$^{4}$He scattering at $E_{\rm lab} = 200$~MeV.  
The right panel indicates that the CF model reproduces $\vec{p}$+$^{6}$He scattering at $E_{\rm lab} = 200$~MeV and that the condition $U_{pn_1}=U_{pn_2}=0$ is good for $d\sigma/d\Omega$ and  $A_y$ 
in $ \theta_{\rm cm} \lsim 52^\circ$. Now we predict $A_y$ for  $\vec{p}$+$^{6}$He scattering at  $E_{\rm lab}=200$~MeV, using the CF model.

Further analyses based on the improved VCC theory are made below by using $q$ instead of $\theta_{\rm cm}$.

\begin{figure}[htbp]
 \begin{center}
  \includegraphics[width=0.9\linewidth]{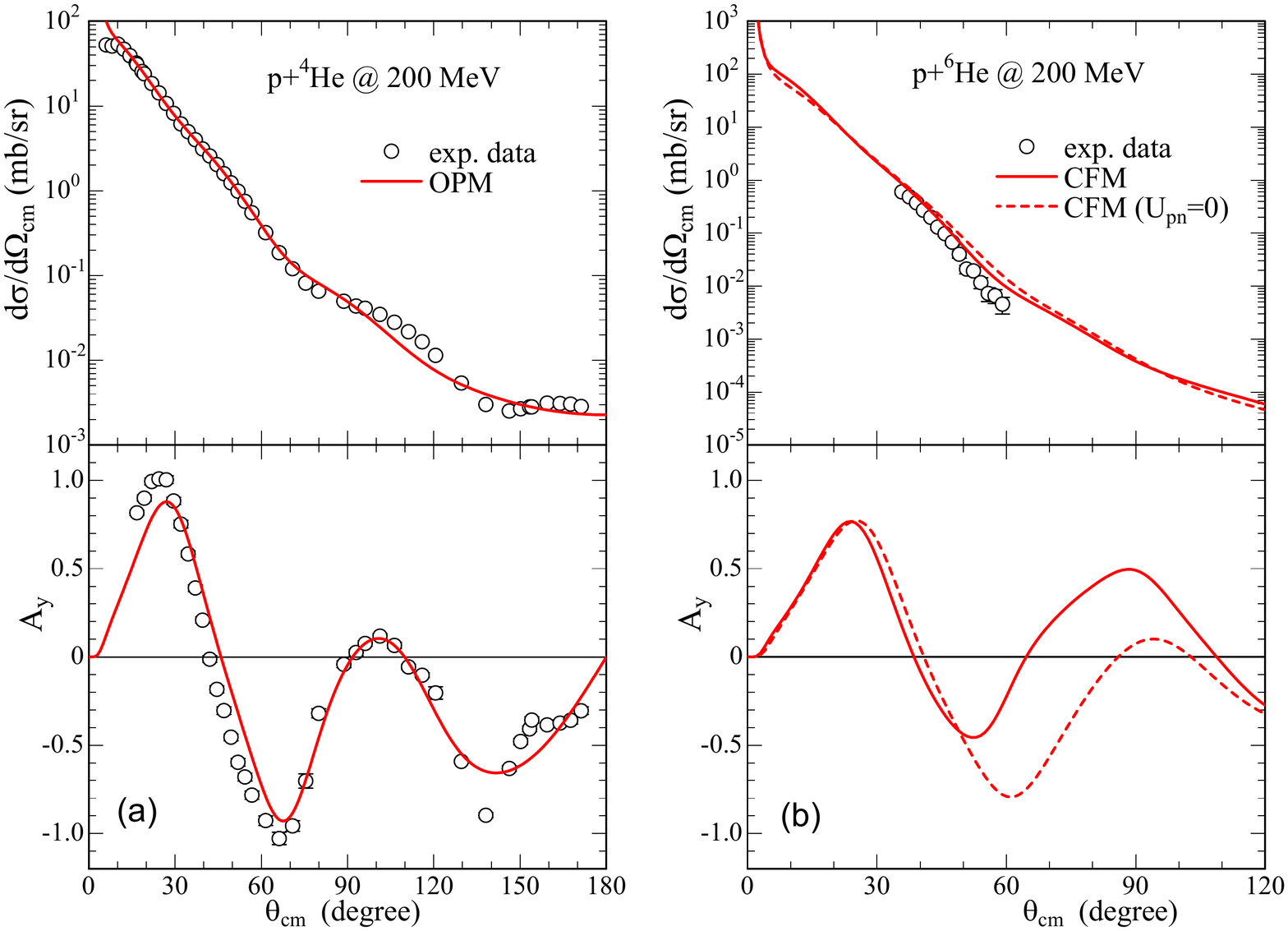}
  \caption{
$\theta_{\rm cm}$ dependence of $d\sigma/d\Omega$ and  $A_y$ for $\vec{p}$+$^{4,6}$He scattering 
at $E_{\rm lab} = 200$~MeV. In left panel, the solid line is a result of our fitting based on the 
optical potential model (OPM).  
In the right panel, the solid and dashed lines denote results of CF model (CFM) 
with and without $U_{pn_1}$ and $U_{pn_2}$, respectively. 
Experimental data are taken from Ref.~\cite{Moss1980} for $^{4}$He and Ref~\cite{Chebotaryov:2018ilv} for $^{6}$He. 
  }
  \label{fig:He4,6-200}
 \end{center}
\end{figure}

\begin{table*}[htbp]
\caption
{Parameters of the optical potentials for $\vec{p}+^4$He scattereing 
at $E_{\rm lab}=200$~MeV. For 72 MeV, the parameter set is taken from Ref.~\cite{Sakaguchi:2011rp}. 
}
\begin{ruledtabular}
\begin{tabular}{ll|cccccccccccc}
 &   &  $V_0$  &  $r_{r}$  &  $a_{r}$  &  $W_0$  &  $r_{i}$ &  $a_{i}$
   &  $W_{id}$  &  $r_{id}$ &  $a_{id}$
   & $V_{s}$ & $r_{s}$ & $a_{s}$ \\
 & (MeV)  &  (MeV)  &  (fm)  &  (fm)  &  (MeV)  &  (fm)  &  (fm)
   &  (MeV)  &  (fm)  &  (fm)
   &   (MeV)  &  (fm)  &  (fm) \\ \hline
$p+^4$He & 200 & -26.528 & 0.7839 & 0.1446 & 17.098 & 1.205 & 0.5268
      & -- & -- & -- & 6.689 & 0.8215 & 0.2641 \\
$p+^4$He & 72 & 54.87 & 0.8566 & 0.09600 & -- & -- & --
      & 31.97 & 1.125 & 0.2811 & 3.925 & 0.8563 & 0.4914 \\
\end{tabular}
\end{ruledtabular}
\label{table:param}
\end{table*}

\subsection{Model-dependent prediction on  $A_y$ for $\vec{p}$+$^{4,6}$He scattering at $E_{\rm lab}=200$~MeV} 
\label{Sec,Xsec,Ay-200}

\begin{figure}[htbp]
 \begin{center}
 \includegraphics[width=0.9\linewidth]{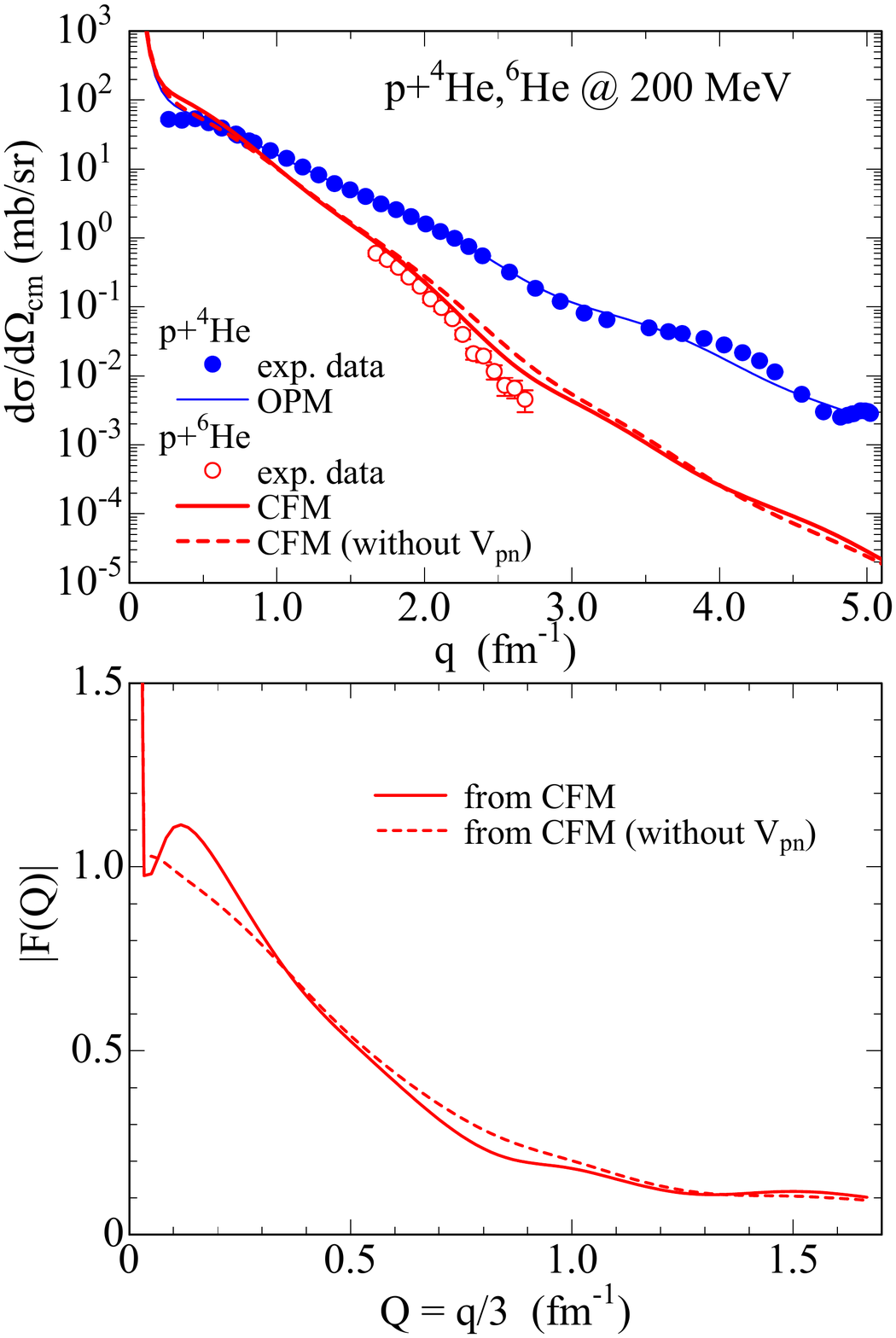}
  \caption{
  $q$ dependence of $d\sigma/d\Omega$ 
for $\vec{p}$+$^{4,6}$He scattering at $E_{\rm lab}=200$~MeV in the upper panel and 
the form factor $|F(Q)|$ in the lower panel. 
Experimental data are taken from Ref.~\cite{Moss1980} for $\vec{p}$+ $^{4}$He scattering
 and Ref.~\cite{Chebotaryov:2018ilv} for $\vec{p}$+$^{6}$He scattering. 
  }
  \label{fig:Xsec-200}
 \end{center}
\end{figure}

Figure~\ref{fig:Xsec-200} shows $q$ dependence of $d\sigma/d\Omega$ 
for $\vec{p}$+$^{4,6}$He scattering at $E_{\rm lab}=200$~MeV in the upper panel and 
the form factor $|F(Q)|$ in the lower panel.
In the upper panel,  the CF model (solid line)  reproduces the data ~\cite{Chebotaryov:2018ilv} for  
$\vec{p}$+$^{6}$He scattering at $E_{\rm lab}=200$~MeV
with no free parameter.
In the lower panel, the solid line denotes the  $|F(Q)|$ calculated with the CF-folding model, while 
$U_{pn_1}$ and $U_{pn_2}$ are switched off in the dashed line. 
The difference between the two lines shows that effects of $U_{pn_1}$ and $U_{pn_2}$ are small 
in the region $0.3 \lsim Q \lsim 0.8$~fm$^{-1}$ ($0.9 \lsim q \lsim 2.4$~fm$^{-1}$).

\begin{figure}[htbp]
 \begin{center}
 \includegraphics[width=0.9\linewidth]{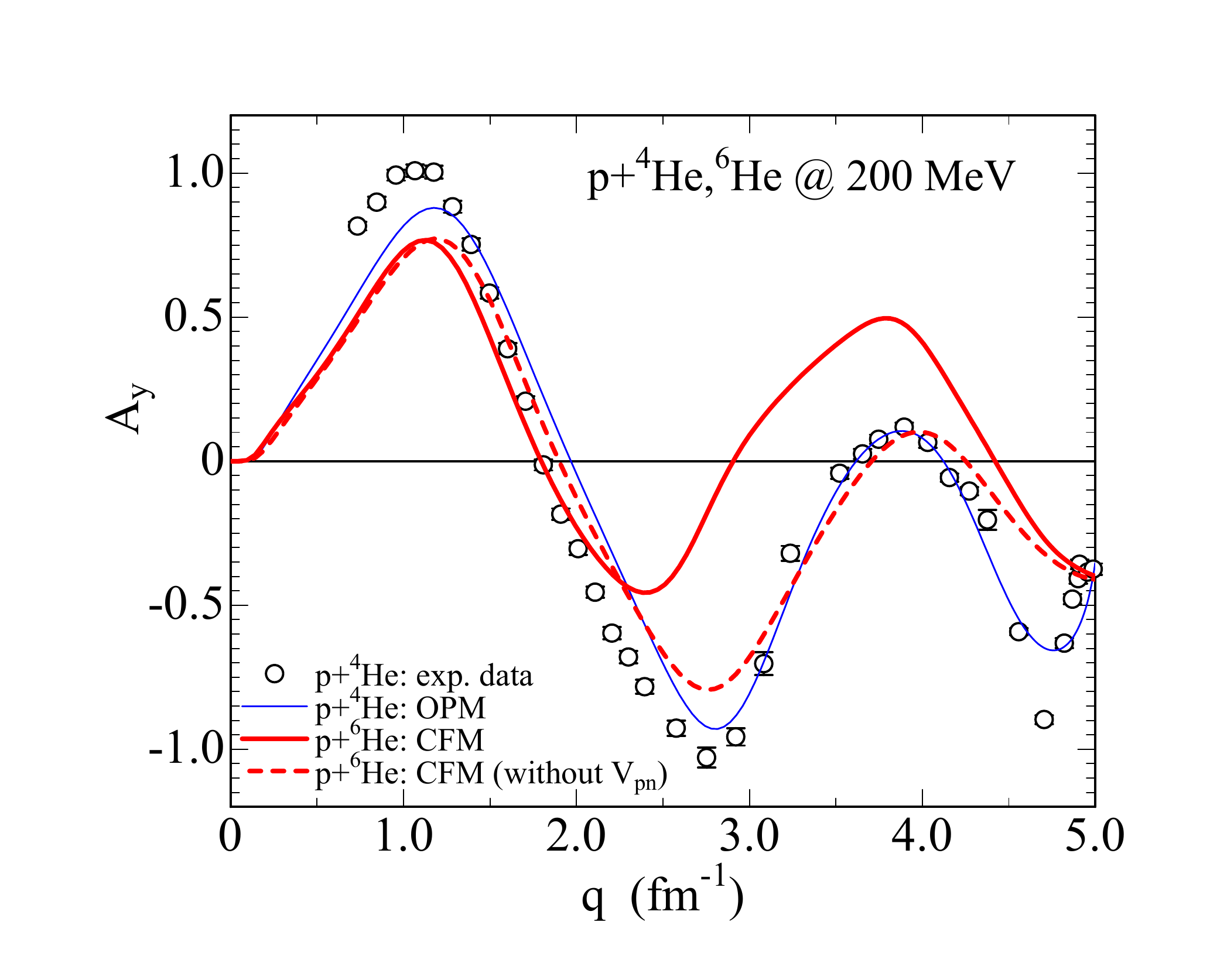} 
   \caption{
  $q$ dependence of $A_y$ for $\vec{p}$+$^{6}$He scattering at $E_{\rm lab}=200$~MeV. 
  The solid line denotes a result of the CF-folding model, while 
  $U_{pn_1}$ and $U_{pn_2}$ are switched off in the dashed line. 
  The thin solid line is a result of the fitting for  $\vec{p}$+$^{4}$He scattering at $E_{\rm lab}=200$~MeV. 
  Open circles show the experimental data~\cite{Moss1980} for $\vec{p}$+ $^{4}$He scattering. 
  In $0.9 \lsim q \lsim 2.4$~fm$^{-1}$, open circles can be regarded as measured $A_y$ for 
  the $\vec{p}$+$^{6}$He scattering at $E_{\rm lab}=200$~MeV. 
  Experimental data are taken from Ref.~\cite{Moss1980} for $^{4}$He.
   }
  \label{fig:Ay-200}
 \end{center}
\end{figure}

Figure~\ref{fig:Ay-200} shows $q$ dependence of $A_y$ for $\vec{p}$+$^{6}$H scattering. 
The solid line denotes the  $A_y$ calculated with the CF-folding model, while
$U_{pn_1}$ and $U_{pn_2}$ are switched off in the dashed line. 
The difference between the solid and dashed lines show that the condition $U_{pn_1}=U_{pn_2}=0$ is good 
in $q \lsim 2.4$~fm$^{-1}$. Eventually, the condition is good in $0.9 \lsim q \lsim 2.4$~fm$^{-1}$, 
when we see both $d\sigma/d\Omega$ and $A_y$.

Now we predict $A_y$ for  $\vec{p}$+$^{6}$He scattering at  $E_{\rm lab}=200$~MeV, using the CF model. 
In $0.9 \lsim q \lsim 2.4$~fm$^{-1}$, open circles are the $A_y$  for $^{6}$He derived from the measured 
$A_y$ of Ref.~\cite{Moss1980} for $^{4}$He. 
The CF model reproduces the derived $A_y$ in $0.9 \lsim q \lsim 2.0$~fm$^{-1}$.

\subsection{CF results on  $d\sigma/d\Omega$ and $A_y$ for 71~MeV} 
\label{Sec,Xsec,Ay-71}

\begin{figure}[htbp]
 \begin{center}
  \includegraphics[width=0.9\linewidth]{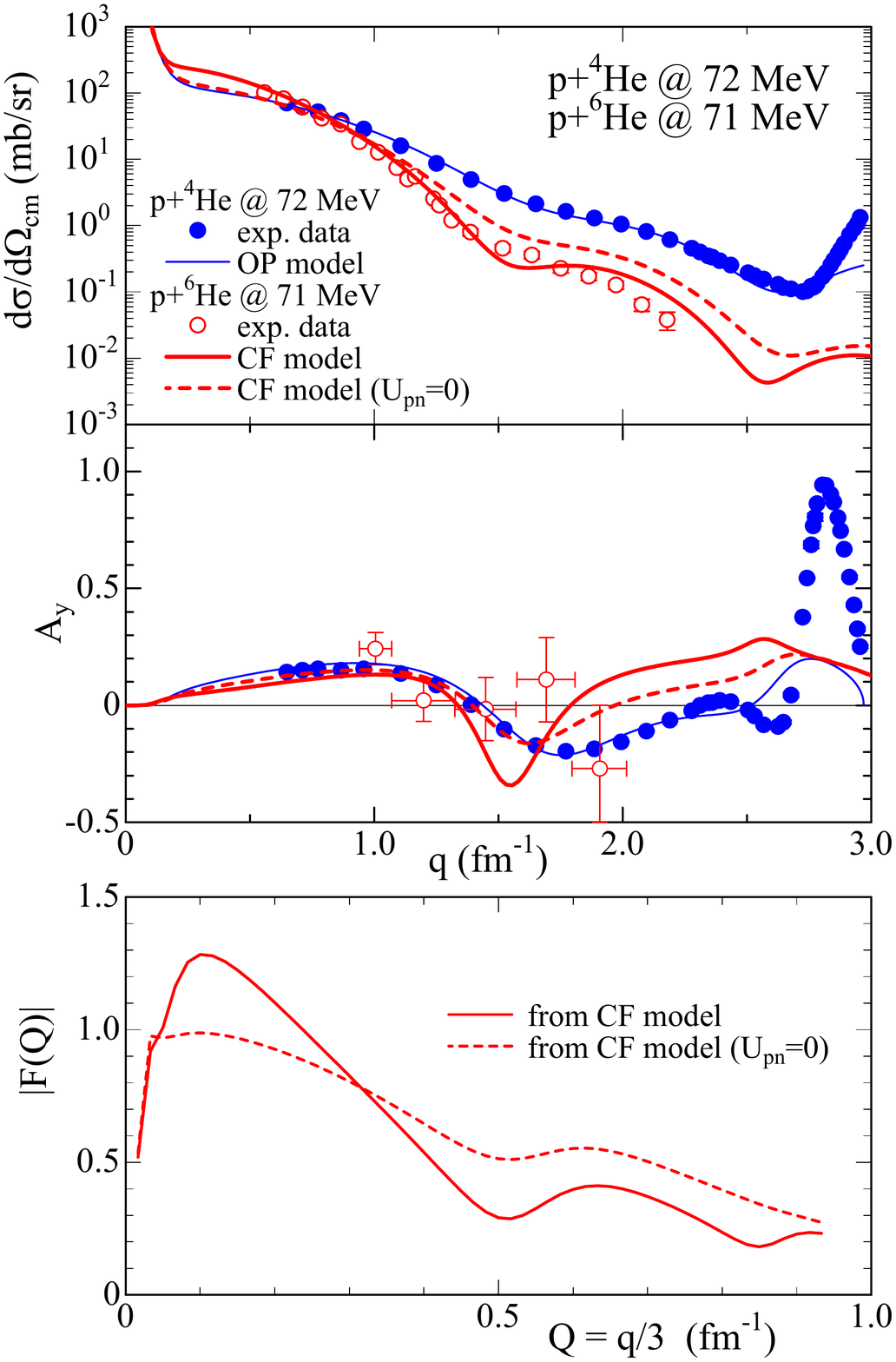}
  \caption{
$q$ dependence of $d\sigma/d\Omega$  and $A_y$ for $\vec{p}$+$^{4,6}$He scattering at $E_{\rm lab} \approx 71$~MeV 
in the upper and middle panels and the form factor $|F(Q)|$ in the lower panel. 
 The solid and dashed lines denote results of CF model with and without $U_{pn_1}$ and $U_{pn_2}$ for $^{,6}$He, respectively. 
 The thin solid line denotes the result of fitting for $^{4}$He. 
 Data are taken from Refs.~\cite{Sakaguchi:2011rp,Korsheninnikov:1997qta} for $^{6}$He  and 
 from Ref.~\cite{Burzynski:1989zz} for $^{4}$He. 
   }
  \label{fig:F-72}
 \end{center}
\end{figure}

Figure~\ref{fig:F-72} shows the results of the CF-model   
for $d\sigma/d\Omega$ and $A_y$ of $\vec{p}$ + $^{6}$He scattering at  $E_{\rm lab} = 71$~MeV 
in the upper and middle panels. The CF model reproduces the data~\cite{Uesaka:2010mm,Sakaguchi:2011rp} with no free parameter. 
The upper and middle panels  also show the results of the best optical potential    
for $d\sigma/d\Omega$ and $A_y$ of $\vec{p}$ + $^{4}$He scattering at  $E_{\rm lab} = 72$~MeV. 

The lower panel shows the $|F(Q)|$ calculated with the CF model.  
The difference between the solid and dashed lines indicates that the condition $U_{pn_1}=U_{pn_2}=0$ 
is good only  in the vicinity of $Q = 0.3$~fm$^{-1}$. 

\section{Summary}
\label{Summary}

We have applied the cluster-folding (CF) model for $\vec{p}+^{6}$He 
scattering at 200~MeV, where the optical potential between $\vec{p}$ and $^{4}$He is fitted to data 
for $\vec{p}+^{4}$He scattering at 200~MeV; see Fig.~\ref{fig:He4,6-200}. 
The CF model reproduces the differential cross section of $\vec{p}+^{6}$He 
scattering with no free parameter.  
We then predict $A_y$, as shown in Fig.~\ref{fig:Ay-200}.
The solid line is our prediction based on the CF model, 
while the open  circles
are our model-independent prediction in $0.9 \lsim q \lsim 2.4$~fm$^{-1}$. 

In order to make the model-independent  prediction for $\vec{p}+^{6}$He scattering at 200~MeV, 
we improve the VCC theory, using the eikonal approximation 
in addition to the $U_{pn_1}=U_{pn_2}=0$ approximation and  the adiabatic approximation. 
In the improved VCC theory, the $A_y$ for  $^{6}$He is the same as that for 
$^{4}$He. 
The $U_{pn_1}=U_{pn_2}=0$ approximation is most essential among the three approximations. 
Using the CF model, we have confirmed 
that the $U_{pn_1}=U_{pn_2}=0$ approximation is good  in $0.9 \lsim q \lsim 2.4$~fm$^{-1}$ 
for 200~MeV, but good only near $q =0.9$~fm$^{-1}$  for 71 MeV. 
In $0.9 \lsim q \lsim 2.4$~fm$^{-1}$, we predict  $A_y(q)$ for  $\vec{p}+^{6}$He scattering at 200~MeV 
from measured $A_y(q)$ for $\vec{p}+^{4}$He scattering at 200~MeV. 
This is a model-independent prediction in $0.9 \lsim q \lsim 2.4$~fm$^{-1}$ 
 ($20^\circ \lsim \theta_{\rm cm} \lsim 55^\circ$); see Fig.~\ref{fig:Ay-exp}.

We thus predict  $A_y(q)$ with the model-dependent and the model-independent prescription. 
Difference between the two predictions is ambiguity of our prediction  in 
$0.9 \lsim q \lsim 2.4$~fm$^{-1}$.

The ratio $|F(Q)|$ of differential cross sections measured for $^{6}$He to that for $^{4}$He 
is related to the wave function of $^{6}$He. We have then determined the radius between  $^{4}$He and 
the center-of-mass of valence two neutrons. The radius is 5.77~fm.

The Jensen-Shannon (JS) divergence~\cite{lin1991divergence} is new data analyses used 
by LIGO Scientific and Virgo Collaborations~\cite{LIGOScientific:2018mvr}. 
The present work is a first application of JS divergence in nuclear physics. 
Since the analysis is two new, we show it in Appendix~\ref{Sec,JSD-Ay-71}.

\noindent
\begin{acknowledgments}
We would like to thank Prof. Hiyama for providing the numerical data of the $^{6}$He density, 
Dr. Toyokawa for providing his code and Prof. Kouno for making comments. 
\end{acknowledgments}

\noindent
\appendix
\section{Jensen-Shannon divergence for $A_y$ for 71~MeV} 
\label{Sec,JSD-Ay-71}

The Jensen-Shannon (JS)  divergence consider two probabilities and make the comparison 
between their shapes quantitatively. We apply the JS divergence to $A_y$ measured for $\vec{p}$+$^{4}$He scattering at $E_{\rm lab} = 72$~MeV and that for $\vec{p}$+$^{6}$He scattering at $E_{\rm lab} = 71$~MeV. 
This is a first application in nuclear physics. For this quantification, we start with two probability distributions, 
$p(q_i)$ and $p(q_i)$, having  $0 \le p(q_i) \le 1$ and $0 \le q(q_i) \le 1$.
The JS divergence is defined 
as~\cite{lin1991divergence} 
\bea
  D_\mathrm{JS} (p||q) =  \sum_{i=1}^{N} D_\mathrm{JS}(q_i)
\label{Eq:JSD}
\eea
with 
\bea
D_\mathrm{JS}(q_i)=
 \frac{1}{2}
 \Bigl[ p(q_i) \ln \Bigl(\frac{p(q_i)}{M(q_i)} \Bigr)
      + q(q_i) \ln \Bigl(\frac{q(q_i)}{M(q_i)} \Bigr)\Bigr],~~ 
\label{Eq:JSD-1}
\eea
for $M(q_i)=(p(q_i)+q(q_i))/2$. The  $D_\mathrm{JS} (p||q)$ 
satifies 
\begin{align}
 D_\mathrm{JS} (p||q) &= D_\mathrm{JS} (q||p),~~~
 0 \le D_\mathrm{JS} (p||q) \le \ln \com{2}=0.693. 
\end{align}
The $D_\mathrm{JS}$ is finite; note that the word   
``divergence" maintains  for historical reasons. 
When the probability distributions are perfectly matched with each other,
the $D_\mathrm{JS}$ becomes exactly zero. The $D_\mathrm{JS}$ becomes $\ln 2=0.693$,
when there are no overlap between the probability distributions. 

In the present data analysis, the number $N$ of data is 5. 
The $\{p_i\}$ are a normalized distribution of measured $(A_y+1)/2$ for $^4$He, while the $\{q_i\}$ are a normalized distribution of measured $(A_y+1)/2$ for $^6$He. The reason why we take $(A_y+1)/2$ is that $0 \le (A_y+1)/2 \le 1$.

Our result $D_\mathrm{JS} \approx 0.0028$ is much smaller than $\ln 2=0.693$. 
This indicates  that the shapes of the two probabilities  are closed to each other. 
The average of $\{p_i\}$ ($\{q_i\}$)  describes the magnitude $M_4$ ($M_6$) for $^4$He ($^6$He).  
The results are $M_4=2.434$ and $M_6=2.539$. The two magnitudes are closed 
to each other, since the difference $(M_6-M_4)/M_6$ is 4 \%. 

When the two  magnitudes are close to each other, we can improve the JS divergence as 
\bea
  D_\mathrm{JS} (p||q) M_{\rm av}=  \sum_{i=1}^{N} D_\mathrm{JS}(q_i) M_{\rm av}
\label{Eq:improved-JSD}
\eea
with 
\bea
&&D_\mathrm{JS}(q_i)M_{\rm av}  
\notag \\       
&\approx& \frac{1}{2}
 \Bigl[ A_y^1(q_i) \ln \Bigl(\frac{2 A_y^1(q_i)}{A_y^1(p_i)+A_y^1(q_i)} \Bigr)
\notag \\       
      &&~~~+ A_y^1(p_i) \ln \Bigl(\frac{2 A_y^1(p_i)}{A_y^1(p_i)+A_y^1(q_i)} \Bigr)\Bigr],~~~~~
\label{Eq: improved-JSD}
\eea
for the average $M_{\rm av}=(M_4+M_6)/2$. and $A_y^1(q_i) \equiv (A_y(q_i)+1)/2$. 
The $D_\mathrm{JS} (p||q) M_a$ describes the magnitude and the shape fo two curves. 
Our result is $D_\mathrm{JS} (p||q) M_a=0.007$ that is much smaller than the maximum $\ln 2*M_a=1.7236$. 
The improved JS divergence thus yields the same conclusion as the original JS divergence. 
The measured $A_y$ for $^{4,6}$He are thus close to each other, although the condition  $U_{pn_1}=U_{pn_2}=0$ 
is not good. There is no theory that explains the similarity.

Now we neglect the data at $q=1.71$~fm$^{-1}$.  The result is $D_\mathrm{JS} (p||q) M_a=0.002$.  
This value is much smaller than $D_\mathrm{JS} (p||q) M_a=0.007$. 
The value of the $D_\mathrm{JS} (p||q) M_a$is much changed by the data at $q=1.71$~fm$^{-1}$. 
We hope that new measurements will be made for $\vec{p}+^{4,6}$He scattering at 71~MeV

\bibliographystyle{prsty}

\begin{thebibliography}{10}



\bibitem{Toyokawa:2013uua} 
  M.~Toyokawa, K.~Minomo and M.~Yahiro,
  Phys.\ Rev.\ C {\bf 88}, no. 5, 054602 (2013).


\bibitem{Watanabe:2014zea} 
  S.~Watanabe {\it et al.},
  Phys.\ Rev.\ C {\bf 89}, no. 4, 044610 (2014).

\bibitem{Hatano2005} 
M. Hatano {\it et al.},
  Eur.\ Phys.\ J.\ A {\bf 25}, 255 (2005).


 
\bibitem{Uesaka:2010mm} 
  T.~Uesaka {\it et al.},
  Phys.\ Rev.\ C {\bf 82}, 021602 (2010),
  [arXiv:1007.3775 [nucl-ex]].
  
\bibitem{Sakaguchi:2011rp} 
  S.~Sakaguchi {\it et al.},
  Phys.\ Rev.\ C {\bf 84}, 024604 (2011),
  [arXiv:1106.3903 [nucl-ex]].
  

\bibitem{Chebotaryov:2018ilv} 
  S.~Chebotaryov {\it et al.},
  PTEP {\bf 2018}, no. 5, 053D01 (2018).
      
  

\bibitem{Moss1980}
G. A. Moss, {\it et~al.}, Phys.\ Rev.\ C{\bf 21},  1932  (1980). 



\bibitem{Burrows:2018ggt} 
  M.~Burrows, C.~Elster, S.~P.~Weppner, K.~D.~Launey, P.~Maris, A.~Nogga and G.~Popa,
  Phys.\ Rev.\ C {\bf 99}, no. 4, 044603 (2019),
  [arXiv:1810.06442 [nucl-th]].
  
\bibitem{Crespo:2007zz} 
  R.~Crespo and A.~M.~Moro,
  Phys.\ Rev.\ C {\bf 76}, 054607 (2007).

\bibitem{Gennari:2017yez} 
  M.~Gennari, M.~Vorabbi, A.~Calci and P.~Navratil,
  Phys.\ Rev.\ C {\bf 97}, no. 3, 034619 (2018). 



\bibitem{Johnson:1997zz} 
  R.~C.~Johnson, J.~S.~Al-Khalili and J.~A.~Tostevin,
  Phys.\ Rev.\ Lett.\  {\bf 79}, 2771 (1997).
  
 \bibitem{Yamaguchi83}
N. Yamaguchi, S. Nagata, and T. Matsuda, Progress of Theoretical Physics {\bf
  70},  459  (1983).

\bibitem{Nagata85}
S. Nagata, M. Kamimura, and N. Yamaguchi, Progress of Theoretical Physics {\bf
  73},  512  (1985).

\bibitem{Yamaguchi86}
N. Yamaguchi, S. Nagata, and J. Michiyama, Progress of Theoretical Physics {\bf
  76},  1289  (1986). 
  
\bibitem{Hamada62}
T. Hamada and I. Johnston, Nuclear Physics {\bf 34},  382  (1962).


\bibitem{Yahiro:2008dr} 
  M.~Yahiro, K.~Minomo, K.~Ogata and M.~Kawai,
  Prog.\ Theor.\ Phys.\  {\bf 120}, 767 (2008).
  [arXiv:0807.3799 [nucl-th]].
  

\bibitem{Korsheninnikov:1997qta} 
  A.~A.~Korsheninnikov {\it et al.},
  Nucl.\ Phys.\ A {\bf 617}, 45 (1997).


\bibitem{Burzynski:1989zz} 
  S.~Burzynski, J.~Campbell, M.~Hammans, R.~Henneck, W.~B.~Lorenzon, M.~A.~Pickar and I.~Sick,
  Phys.\ Rev.\ C {\bf 39}, 56 (1989).

 
  



\bibitem{Hiyama96}
E. Hiyama {\it et~al.}, Physical Review C {\bf 53},  2075  (1996).

\bibitem{Hiyama03}
E. Hiyama, Y. Kino, and M. Kamimura, Progress in Particle and Nuclear Physics
  {\bf 51},  223  (2003).





  


  
\bibitem{lin1991divergence}
J.~Lin, IEEE Transactions on Information theory, {\bf 37}, 145 (1991). 

\bibitem{LIGOScientific:2018mvr} 
  B.~P.~Abbott {\it et al.} [LIGO Scientific and Virgo Collaborations],
  Phys.\ Rev.\ X {\bf 9}, no. 3, 031040 (2019),
  [arXiv:1811.12907 [astro-ph.HE]].



\end{thebibliography}

\end{document}